\documentclass[12pt]{article}
\usepackage{a4wide}
\usepackage{amssymb}
\usepackage{amsmath}
\usepackage{mathrsfs}
\usepackage{graphicx}
\usepackage{caption}
\usepackage{subcaption}
\begin{document}
{\renewcommand{\thefootnote}{\fnsymbol{footnote}}
\medskip
\begin{center}
{\Large  Infinite circuits are easy. How about long ones?}\\
\vspace{1.5em}
Mikhail Kagan, Xinzhe Wang\\
Penn State Abington
\vspace{1.5em}
\end{center}
}

\setcounter{footnote}{0}
\newcommand{\Lam}{\Lambda}
\newcommand{\vt}{\vartheta}
\newcommand{\be}{\begin{equation}}
\newcommand{\ee}{\end{equation}} 
\newcommand{\bq}{\begin{eqnarray}} 
\newcommand{\eq}{\end{eqnarray}}
\newcommand{\f}{\frac}
\newcommand{\vp}{\varphi}
\newcommand{\abs}[1]{\lvert#1\rvert}

\newcommand{\case}[2]{{\textstyle \frac{#1}{#2}}}
\newcommand{\lP}{\ell_{\mathrm P}}

\newcommand{\md}{{\mathrm{d}}}
\newcommand{\Kern}{\mathop{\mathrm{ker}}}
\newcommand{\tr}{\mathop{\mathrm{tr}}}
\newcommand{\sgn}{\mathop{\mathrm{sgn}}\nolimits}

\newcommand*{\R}{{\mathbb R}}
\newcommand*{\N}{{\mathbb N}}
\newcommand*{\Z}{{\mathbb Z}}
\newcommand*{\Q}{{\mathbb Q}}
\newcommand*{\C}{{\mathbb C}}

\begin{abstract}
We consider a long but finite (ladder) circuit with alternating connections of resistors in series and parallel and derive an explicit expression for its equivalent resistance as a function of the number of repeating blocks, $R_{\rm eq}(k)$. This expression provides an insight on some {\em adjacent} topics, such as continued fractions and a druncard\rq{}s random walk in a street with a gutter. We also remark on a possible method of solving the non-linear recurrent relation between $R_{\rm eq}(k)$ and $R_{\rm eq}(k+1)$.  The paper should be accessible to students familiar with the arithmetics of determinants. 
\end{abstract}

\section*{Introduction}
\begin{figure}[h!] 
\centerline{\includegraphics[width=10cm, keepaspectratio]{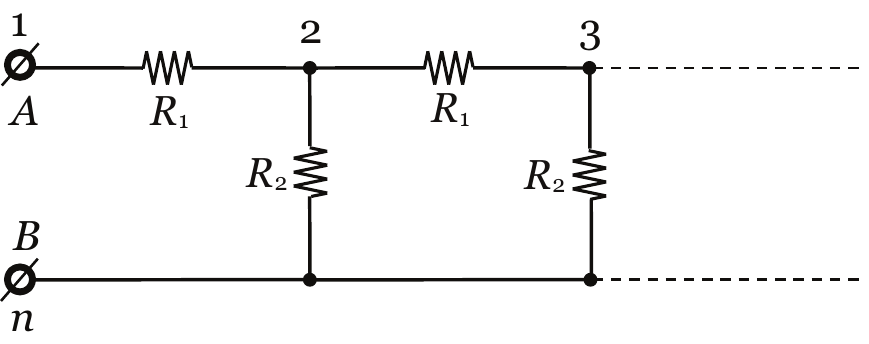}} \caption{The classical infinite circuit composed of alternative parallel and series connections of resistors $R_1$ and $R_2$.  \label{Fig:InfCirc}} 
\end{figure}

A well known classical problem about finding the equivalent resistance of an inifinite electric circuit may quickly become much more interesting if the circuit is no longer infinite. For example, the infinite circuit in Fig. \ref{Fig:InfCirc}, also known as a {\em ladder} circuit, with all resistors equal ($R_1=R_2=R$) is known to have the \lq{}\lq{}golden ratio\rq{}\rq{} equivalent resistance
\be\label{Golden_Ratio}
R_{\rm eq} = \frac{\sqrt{5}+1}{2}R
\ee
across points $A$ and $B$. This solution utilizes the fact that cutting one (or more) repeating block(s) out of the circuit does not affect the equivalent resistance. This idea clearly would not work if one is dealing with a long but finite circuit, say with a hundred or thousand repeating $R_1$\rq{}s and $R_2$\rq{}s.
Finite ladder circuits have always been popular among physics instructors as a source of interesting laboratory activities \cite{Babylonian,Irrational} and for their tight links with many interesting aspects of number theory, such as continued fractions, golden ratio and the Fibonacci numbers \cite{Irrational,Fibonacci}.
The goal of this work is to obtain a generic closed formula for the equivalent resistance of the circuit from Fig. \ref{Fig:InfCirc} with precisely $k$ repeating blocks, $R_{\rm eq}(k)$.

The paper is organized as follows. In Section \ref{Sec:InfCirc}, we review the classical problem of computing the equivalent resistance of an infinite circuit from Fig. \ref{Fig:InfCirc}. Then we consider the corresponding finite circuit in Section \ref{Sec:FinCirc}, which we start by briefly reviewing the closed formula for the equivalent resistance of a generic circuit recently derived in Ref. \cite{Req}. In Sections \ref{Sec:Determinants} and \ref{Sec:General} we derive the closed formulas for the equivalent resistance as a function of the number of repeating blocks, $R_{\rm eq}(k)$. In Section \ref{Sec:Discussion}, we remark on the correspondence between this electric circuit and continued fractions and present a method of solving the arising non-linear recurrent relation between $R_{\rm eq}(k)$ and $R_{\rm eq}(k+1)$. At the end we discuss the analogy between the ladder sircuit and a druncard\rq{}s random walk in a street with a gutter.

\section{Review of an Infinite Circuit}\label{Sec:InfCirc}
\begin{figure}[h]
        \centering
        \begin{subfigure}[b]{0.45\textwidth}
                \centering
                \frame{\includegraphics[width=\textwidth]{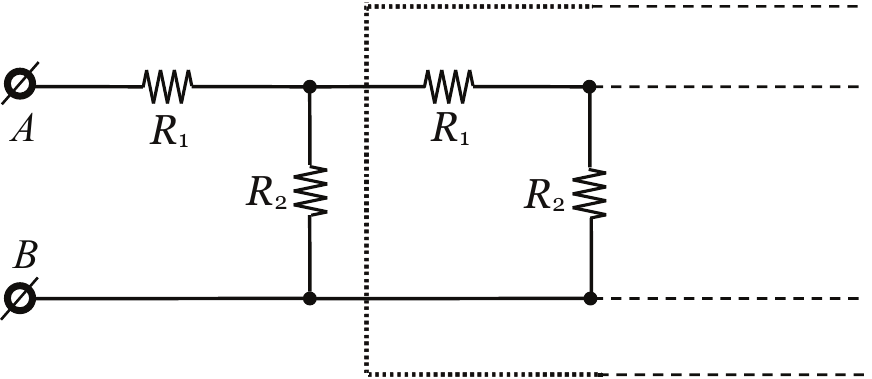}}
                \caption{The circuit in the box has the same resistance as the original circuit.}
                \label{Fig:InfCircBox}
        \end{subfigure}%
        \qquad
        \begin{subfigure}[b]{0.40\textwidth}
                \centering
                \frame{\includegraphics[width=\textwidth]{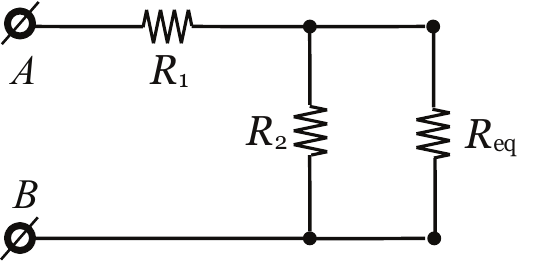}}
                \caption{The boxed part of the circuit from Fig.\ref{Fig:InfCircBox} is connected in parallel with $R_2$.}
                \label{Fig:InfCircPar}
        \end{subfigure}%
		\quad
         \caption{Classical solution of the infinite circuit problem.}
\end{figure}
The infinite circuit shown in Fig. \ref{Fig:InfCirc} is assembled from two sets of identical resistors, $R_1$ and $R_2$, with alternating connections in series and in parallel. Since the former type of connection increases the overall resistance and the latter one dereases, the equivalent resistance of the entire circuit is finite and can be computed using the folowing idea. As the circuit is infinite, cutting one (or more) repeating block(s) off the circuit yields exactly the same circuit. In other words, the equivalent resistance of the boxed part of the circuit (see Fig. \ref{Fig:InfCircBox}) is equal to that of the original circuit.  Thus this boxed part has resistance $R_{\rm eq}$ and is connected in parallel with the leftmost resistor $R_2$ (see Fig. \ref{Fig:InfCircPar}). Expressing the overal resistance of this three-resistor circuit and setting it equal to $R_{\rm eq}$, we obtain
\be\label{ResRecursion}
R_{\rm 1}+\left(\frac{1}{R_{\rm 2}}+\frac{1}{R_{\rm eq}}\right)^{-1} =R_{\rm eq}.
\ee
Solving the resulting quadratic equation for $R_{\rm eq}$ and picking the positive root, we obtain
\be\label{R_Inf}
R_{\rm eq}=\frac{R_1+\sqrt{R_1^2+4R_1R_2}}{2}.
\ee
In particular, for $R_{1}=R_{2}=R$, we recover the famous \lq\lq{}golden ratio\rq\rq{} answer $R_{\rm eq}=\frac{1+\sqrt{5}}{2}R$ in Eq. (\ref{Golden_Ratio}).

\section{Finite Circuit}\label{Sec:FinCirc}
Let us now consider a finite circuit with $k$ repeating units (pairs of $R_1$ and $R_2$). Such a circuit would have $n=k+2$ nodes (junctions), as illustrated in Fig. \ref{Fig:InfCirc}. Nodes $A$ and $B$ are labeled with $1$ and $n$ respectively. The minimum value of $n$ is $3$ for one pair of resistors. Then each new pair adds one more node. Note that the entire bottom wire is at the same potential equal to that at $B$, hence all the bottom nodes can be identified with $B$. Below we consider an unsuccessful (Section \ref{Sec:Recursion}) and sucessful (Section \ref{Sec:Determinants}) attempts to compute the equivalent resistance/conductance of this circuit.
\subsection{Why straightforward recursive approach does not work}\label{Sec:Recursion}
It is tempting to try mathematical induction, that is to start with a small circuit (few repeating blocks) and add one pair of $R_1$ and $R_2$. In that one immediately faces the following dilemma: the new pair can be added to the left or to the right from the previous circuit. These two operations lead to different outcomes and only one of them is actually correct. Indeed, adding two new resistors to the right of the existing circuit would violate the assumption that the rightmost $R_1$ and $R_2$ of that circuit were in series. Besides, adding more and more blocks in this manner would always increase the number of {\em parallel} connections, thus leading to a {\em zero} overall resistance in the limit, which is clearly incompatible with Eq. (\ref{Golden_Ratio}).

  Adding the new pair of resistors to left of the previous circuit (with equivalent resistance $R_{\rm eq}^{(n)}$) can be illustrated by the same Figure \ref{Fig:InfCircBox}. Repeating the calculation similar to Eq. (\ref{ResRecursion}), we obtain the following recursion relation for the equivalent resistance
\bq
R_{\rm eq}^{(n+1)} &=& R_{1}+\left(\frac{1}{R_{2}}+\frac{1}{R_{\rm eq}^{(n)}}\right)^{-1} \nonumber\\
&\equiv&R_1+\frac{R_{\rm eq}^{(n)} R_2}{R_{\rm eq}^{(n)}+R_2}.\label{Recursion}
\eq
Prior to the writing of this paper, we were unaware of a regular method of solving such a non-linear functional equation, even if $R_1=R_2=R$. Now, in the hindsight, we do. A possible method is presented in Section \ref{Rec_Soln}.
\subsection{Closed formula for the equivalent resistance of a generic circuit}
\begin{figure}[h!] 
\centerline{\includegraphics[width=10cm, keepaspectratio]{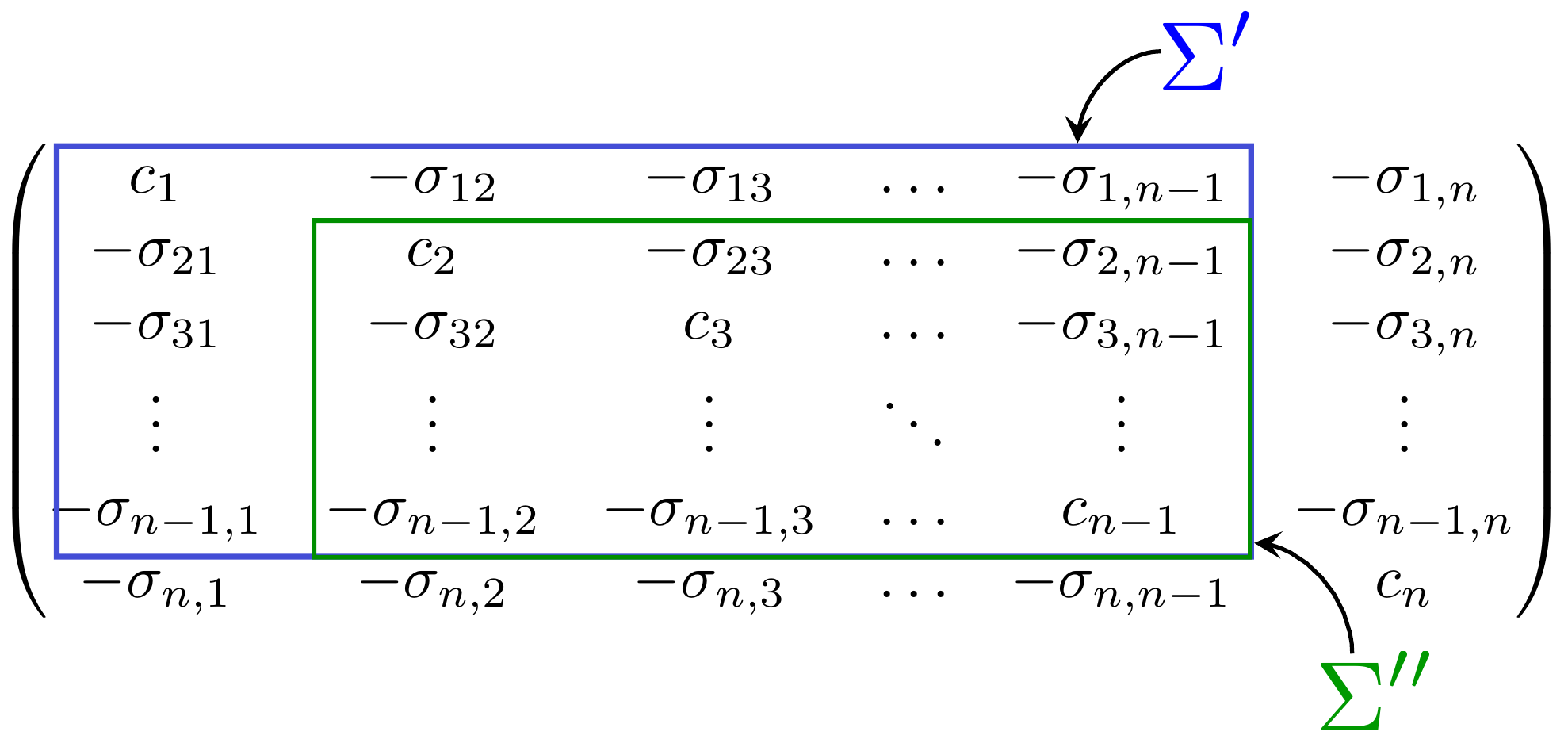}} \caption{Construction of the $\Sigma$-matrix. Sub-matrices $\Sigma^\prime$ and $\Sigma^{\prime \prime}$ are used to compute the equivalent conductance/resistance of the circuit. \label{Fig:SigmaMatrix}} 
\end{figure}

It turns out that it is possible to find the equivalent resistance of the finite circuit in Fig. \ref{Fig:InfCirc} using the approach presented in Ref. \cite{Req}. For convenience, we shall work with conductance (inverse resistance) rather than resistance. For a circuit with $n$ nodes, designate the edge conductance between nodes $i$ and $j$ as $\sigma_{ij}=1/R_{ij}=\sigma_{ji}\geq 0$ and fill out the  $n\times n$ matrix $\Sigma$ in Fig. \ref{Fig:SigmaMatrix}.
There, the diagonal entries are the \lq\lq{}total\rq\rq{} conductance at the corresponding node 
\be\label{c_i}
c_{i}=\sum\limits_{j=1}^n \sigma_{ij}.
\ee
 Assuming that the battery is connected to the first and last nodes, we will need two sub-matrices $\Sigma^\prime$ and $\Sigma^{\prime\prime}$ (see Fig. \ref{Fig:SigmaMatrix}), in terms of which the equivalent conductance of the circuit is given by 
\be\label{Ceq}
\sigma_{\rm eq}=\frac{\det \Sigma^\prime}{\det \Sigma^{\prime\prime}}.
\ee
Below we apply this formula to compute the equivalent conductance of the circuit in Fig. \ref{Fig:InfCirc}.
\subsection{Derivation of the equivalent resistance of a finite circuit}\label{Sec:Determinants}
Adopting the notation introduced in the beginning of Section \ref{Sec:FinCirc}, we can construct the relevant $\Sigma$-matrix. Note that node $1$ is only connected to node $2$; node $n$ is connected to every node except the $1^{\rm st}$ one; and a generic node $k$ ($k\neq 1$ or $n$) is connected to nodes $k-1$, $k+1$ and $n$. Finally, setting $R_1=R_2=1\Omega$ ($\sigma_1=\sigma_2=1 \Omega^{-1}$), we write
\be\label{SigmaMatrix}
\Sigma=\begin{pmatrix}
1 & -1 & 0 & 0 & ...&0& 0  &0 \\
-1 & 3 & -1 & 0  &... &0& 0 &-1\\
0 & -1 & 3 & -1 &  ...&0& 0  &-1\\
0 & 0 & -1 & 3  & ...&0& 0 &-1 \\
\vdots&\vdots &  \vdots & \vdots&\ddots &\vdots & \vdots&\vdots \\
0 & 0 & 0 & 0 &  ... &3& -1  &-1\\
0& 0 & 0 & 0 &...&-1 & 2 & -1 \\
0& -1 & -1 & -1 &... &-1& -1 & n-2 
\end{pmatrix}
\ee
For the determinants , we get 
\be\label{DeltaPrime}
\det\Sigma^{\prime}=\begin{vmatrix}
1 & -1 & 0 &  ...&0& 0   \\
-1 & 3 & -1 &... &0& 0 \\
0 & -1 & 3  &  ...&0& 0  \\
\vdots&\vdots &  \vdots &\ddots &\vdots & \vdots&\\
0 & 0 & 0  &  ... &3& -1  \\
0& 0 & 0 &...&-1 & 2 & \\
\end{vmatrix},
\quad
\det\Sigma^{\prime\prime}=\begin{vmatrix}
 3 & -1 &... &0& 0 \\
 -1 & 3  &  ...&0& 0  \\
\vdots &  \vdots &\ddots &\vdots & \vdots&\\
 0 & 0  &  ... &3& -1  \\
 0 & 0 &...&-1 & 2 & \\
\end{vmatrix},
\ee
of  size $(n-1)\times(n-1)$ and $(n-2)\times(n-2)$ respectively.
Expanding the  $\det \Sigma^\prime$ in the elements of the first row, we obtain
\be\label{sum1}
\det\Sigma^{\prime}=\det\Sigma^{\prime\prime}+\begin{vmatrix}
-1 &  -1 &... &0& 0 \\
0 &  3  &  ...&0& 0  \\
\vdots &  \vdots &\ddots &\vdots & \vdots&\\
0 & 0   &  ... &3& -1  \\
0& 0 &...&-1 & 2 & \\
\end{vmatrix}.
\ee
The last determinant can be written as $-\det\Sigma^{\prime\prime\prime}$, where the matrix $\Sigma^{\prime\prime\prime}$ has the same structure as $\Sigma^{\prime\prime}$, but is $(n-3)\times(n-3)$ in size. More specifically, the matrices of interest have a $2$ in the bottom right corner, $3$\rq{}s everywhere else on the main diagonal, $-1$\rq{}s on the  two adjacent diagonals and zero elsewhere. It will be convenient to denote determinants of these matrices $\Delta_k$, such that
\be
\det\Sigma^{\prime\prime}=\Delta_k, \quad \det\Sigma^{\prime\prime\prime}=\Delta_{k-1},
\ee
where $k$ denotes the size of the matrix and runs from $n-2$ down to $1$. For example, 
\be\label{InCond}
\Delta_1=\begin{vmatrix}
2
\end{vmatrix}=2, \quad \Delta_2 = \begin{vmatrix}
3 &  -1  \\
-1 &  2   \\
\end{vmatrix}=5, \quad \Delta_3= \begin{vmatrix}
3 &  -1 &0 \\
-1&3&-1\\
0&-1 &  2   \\
\end{vmatrix}=13.
\ee
Using this notation in Eq. (\ref{sum1}), we can write 
\be \label{DeltaRec}
\det \Sigma^{\prime}=\Delta_{k}-\Delta_{k-1}
\ee
(with $k=n-2$) and substitute it into Eq. (\ref{Ceq}), which results in
\be\label{Req}
R_{\rm eq}\equiv \frac{1}{\sigma_{\rm eq}}=\frac{\det \Sigma^{\prime\prime}}{\det\Sigma^{\prime}}=\frac{\Delta_{k}}{\Delta_{k}-\Delta_{k-1}}\equiv\frac{1}{1-\frac{\Delta_{k-1}}{\Delta_{k}}}.
\ee
Thus knowing the expression of the $k$-th determinant as an explicit function of $k$ would yield the desired formula for the equivalent resistance of the finite circuit in Fig. \ref{Fig:InfCirc}. Below we find such a formula by obtaining a recurrent relation for $\Delta_k$ and then solving it.

First, we expand  the determinant $\det\Sigma^{\prime\prime}$ from Eq.(\ref{DeltaPrime}) in the elements of the first row
\bq
\Delta_k&=&3\begin{vmatrix}
 3 & -1 &... &0& 0 \\
 -1 & 3  &  ...&0& 0  \\
\vdots &  \vdots &\ddots &\vdots & \vdots&\\
 0 & 0  &  ... &3& -1  \\
 0 & 0 &...&-1 & 2 & \\
\end{vmatrix}
+\begin{vmatrix}
 -1 & -1 &... &0& 0 \\
 0 & 3  &  ...&0& 0  \\
\vdots &  \vdots &\ddots &\vdots & \vdots&\\
 0 & 0  &  ... &3& -1  \\
 0 & 0 &...&-1 & 2 & \\
\end{vmatrix} \nonumber\\
&=&3\Delta_{k-1}-\Delta_{k-2},\label{RecRn}
\eq
where the last determinant was expanded by the elements of its first column. This recurrent relation is linear and very well studied. Much like in linear differential equations, the solution(s) is sought in the form of a geometric progression $x^k$, with $x$ being a constant. Substituting this ansatz into Eq.(\ref{RecRn}) results in $x^{k}=3x^{k-1}-x^{k-2}$, which is equivalent to a quadratic equation in $x$
\be
 x^{2}-3x+1=0, 
\ee
with two solutions 
\be\label{Roots}
x_{1,2}=\frac{3\pm\sqrt{5}}{2}\equiv \left(\frac{1\pm\sqrt{5}}{2}\right)^2.
\ee
The latter form is another appearance of the golden ratio. The general solution of Eq. (\ref{RecRn}) is a linear combination 
\be\label{a_k}
\Delta_{k}=A_{1}x_{1}^{k}+A_{2}x_{2}^{k}, 
\ee
where the constants $A_{1,2}$ are to be determined from the initial conditions (\ref{InCond}). Before we do so, we would like to make the following remark.
The second root in Eq.(\ref{Roots}) $x_{2}=\frac{3-\sqrt{5}}{2}<1$, so as $k$ goes infinity, the term of $A_{2}x_{2}^{n}$ in Eq. (\ref{a_k}) can be neglected.  Hence the ratio of two consecutive determinants 
\be
\frac{\Delta_{k}}{\Delta_{k-1}}\rightarrow x_1 = \frac{3+\sqrt{5}}{2},\quad {\rm as} \quad k\rightarrow \infty.
\ee
Substituting this ratio into Eq. (\ref{Req}) yields $R_{\rm eq}=\frac{1+\sqrt{5}}{2}$, which is the answer we saw earlier in the end of Section \ref{Sec:InfCirc}.
Let us now determine the coefficients $A_1$ and $A_2$ by substituting $k=1$ and $k=2$ into Eq.(\ref{a_k}) and equating the corresponding expressions to the initial conditions in Eq. (\ref{InCond}). This leads to the following system of two equations
\bq
a_{1}&=&A_{1}x_{1}+A_{2}x_{2}=2, \nonumber\\ a_{2}&=&A_{1}x_{1}^{2}+A_{2}x_{2}^{2}=5, 
\eq
whose solution is $A_{1,2}=\frac{\sqrt{5}\pm 1}{2\sqrt{5}}$. Therefore, the $k^{\rm th}$ determinant  in Eq. (\ref{a_k}) reads

\be
\Delta_k=\frac{1}{\sqrt{5}}\left[\left(\frac{\sqrt{5}+1}{2}\right)^{2k+1}+\left(\frac{\sqrt{5}-1}{2}\right)^{2k+1}\right],
\ee
which can be recognized as the odd-numbered Fibonacci number, $F_{2k+1}$. Substituting this expression in Eq. (\ref{Req}) and using the fact that $F_{k+1}=F_{k}+F_{k-1}$, yields
\be\label{REQ}
R_{\rm eq}(k) = \frac{F_{2k+1}}{F_{2k+1}-F_{2k-1}}\equiv \frac{F_{2k+1}}{F_{2k}},
\ee
e.g., a ratio of two consecutive Fibonacci\rq{}s. Here $k=1, 2, 3, \dots$.
\subsection{More general circuit}\label{Sec:General}

In the previous section, we discussed a special case where all resistors have resistance 1$\Omega$. In this section, we consider a more general circuit from Fig. \ref{Fig:InfCirc}, such that all resistors on the horizontal wire have resistance $R_{1}$ (conductance $\sigma_{1}$) and all resistors on the vertical wire have resistance $R_{2}$ (conductance $\sigma_{2}$). The $\Sigma$-matrix from Eq.(\ref{SigmaMatrix})  then takes the form

\be\label{DeltaPrime_General}
\Sigma=\begin{pmatrix}
\sigma_1 & -\sigma_1 & 0 &  ...&0& 0 &0  \\
-\sigma_1& 2\sigma_1+\sigma_2 & -\sigma_1 &... &0& 0 & -\sigma_2\\
0 & -\sigma_1 & 2\sigma_1+\sigma_2  &  ...&0& 0 &-\sigma_2 \\
\vdots&\vdots &  \vdots &\ddots &\vdots & \vdots& \vdots\\
0 & 0 & 0  &  ... &2\sigma_1+\sigma_2& -\sigma_1 &-\sigma_2 \\
0& 0 & 0 &...&-\sigma_1 & \sigma_1+\sigma_2 & -\sigma_2\\
0& -\sigma_2& -\sigma_2 &...&-\sigma_2 &-\sigma_2 & \sigma_2(n-2)
\end{pmatrix}.
\ee
Below we highlight the changes in the main equations. Firstly, instead of the expansion (\ref{DeltaRec}) of $\det \Sigma^\prime$ we have
\be \label{DeltaRec_General}
\det \Sigma^{\prime}=\sigma_1\Delta_{k}-\sigma_1^2\Delta_{k-1},
\ee
which leads to
\be\label{Req_General}
R_{\rm eq}=\frac{\det \Sigma^{\prime\prime}}{\det\Sigma^{\prime}}=\frac{\Delta_{k}}{\sigma_1\Delta_{k}-\sigma_1^2\Delta_{k-1}}\equiv\frac{1/\sigma_1}{1-\frac{\sigma_1\Delta_{k-1}}{\Delta_{k}}}.
\ee
The new recurrence relation for the determinants becomes
\be
\Delta_{k}=(2\sigma_{1}+\sigma_{2})\Delta_{k-1}-\sigma_1^2 \Delta_{k-2}.\label{RecRn_General}
\ee
Its solution is now given by
\be
\Delta_k=\frac{\left(4\sigma_1+\sigma_2\right)^k}{2^{2k+1}}\left[\left({\sqrt{\frac{\sigma_2}{4\sigma_1+\sigma_2}}+1}\right)^{2k+1}-\left({\sqrt{\frac{\sigma_2}{4\sigma_1+\sigma_2}}-1}\right)^{2k+1}\right].
\ee
Substituting this expression into Eq. (\ref{Req_General}) and putting $1/\sigma_1\equiv R_1=R$, $1/\sigma_2 \equiv R_2=\alpha R$, we obtain the general expression for the equivalent resistance
\be\label{REQ}
R_{\rm eq}(k)=\frac{R}{2}\cdot\frac{\left({1+\sqrt{{1+4\alpha}}}\right)^{2k+1}-\left({1-\sqrt{{1+4\alpha}}}\right)^{2k+1}}{\left({1+\sqrt{{1+4\alpha}}}\right)^{2k}-\left({1-\sqrt{{1+4\alpha}}}\right)^{2k}}.
\ee
Note that for $k\rightarrow \infty$, the first terms in both numerator and denominator in Eq. (\ref{REQ}) dominate and the equivalent resistance approaches 
\be
\frac{R}{2}\left(1+\sqrt{1+4\alpha}\right),
\ee
which gives the golden ratio result for identical resistors ($\alpha=1$). It is also easy to see that Eq. (\ref{REQ}) has the right behavior when $\alpha \rightarrow 0$. Physically, such a limit corresponds to short-circuiting all the resistors in the circuit but the very first $R_1$ (=$R$). Thus $R_{\rm eq} \rightarrow R$, which is consistent with Eq. (\ref{REQ}) for $\alpha=0$.
\section{Discussion}\label{Sec:Discussion}
In this section, we discuss some interesting implications of the equivalent resistance formula (\ref{REQ}).
\subsection{Equivalent resistance as a continued fraction}
One \lq{}\rq{}straightforward\rq{}\rq{} way of computing the equivalent resistance of the circuit in Fig. \ref{Fig:InfCirc} is alternating the basic rules for resistors in series and in parallel and writing $R_{\rm eq}$ as a continued fraction
\be\label{ContFrac}
R_{\rm eq} = R_1 + \frac{1}{\frac{1}{R_2}+\frac{1}{R_1+\frac{1}{\frac{1}{R_2}+\frac{1}{R_1+\dots}}}} \equiv R \left( 1 + \frac{\alpha}{1+\frac{\alpha}{1+\frac{\alpha}{1+\frac{\alpha}{1+\dots}}}} \right),
\ee
where, as before, $R=R_1$ and $\alpha = R_2/R_1$. Therefore, we can conclude that the continued fraction above is equivalent to Eq. (\ref{REQ}).
\subsection{Ad-hoc solution to the recursion relation in Eq. (\ref{Recursion})}\label{Rec_Soln}
In the hindsight, we can explicitly solve the recursion relation in Eq. (\ref{Recursion}) using the following ansatz
\be\label{Ansatz}
R_{\rm eq}^n = \frac{S_{n+1}}{S_n},
\ee
with some undetermined sequence $S_n$. Substituting this ansatz into Eq. (\ref{Recursion}) yields 
\be
S_{n+2}S_{n+1}+R_2 S_{n+2} S_n - (R_1+R_2)S_{n+1}^2 - R_1 R_2 S_{n+1} S_n =0,
\ee
which can be solved using the substitution $S_n = x^n$. It is easy to see that $x$  satisfies
\be
x^2-R_1 x - R_1 R_2 = 0,
\ee
whose solution is $x_{1,2}=\frac{R_1\pm\sqrt{R_1^2+4R_1R_2}}{2}$. Note that the larger root corresponds to the equivalent resistance of an infinite circuit in Eq. (\ref{R_Inf}). The general solution for $S_n$  is given by a linear combination $A_1 x_1^n + A_2 x_2^n$. Clearly, using the initial conditions in Eq. (\ref{InCond}) and substituting $S_n$ in Eq. (\ref{Ansatz}) results in Eq. (\ref{REQ}).
\subsection{Equivalent resistance and a random walk}
There is an elegant analogy between electrical circuits and random walks on graphs \cite{Doyle}. The latter are described by the {\em transition probability} matrix, $P_{ij}$, that specifies the probability of going from one node of the graph ($i$) to another one ($j$). This matrix can be obtained from the conductance matrix by the following ``normalization\rq{}\rq{}
\be
P_{ij}=\frac{\sigma_{ij}}{c_i}, 
\ee
with $c_i$ given by Eq. (\ref{c_i}). This definition ensures that $P_{ij}\geq 0$ and $\sum\limits_j{P_{ij}}=1$.  In the special case of all edges having the same resistance, such probability is the inverse of the number of edges coming out of the first node ($i$). Note that in general, the matrix $P_{ij}$ is not symmetric. The direct analog of the equvalent conductance is so called {\em escape probability}, $P_{xy}^{\rm esc}$, that is the probabilty of a random walk that started at node $x$ to reach node $y$ before coming back to $x$. The relationship between the equivalent conductance of an electrical circuit (between nodes $x$ and $y$) and the escape probability of the corresponding graph is given by
\be\label{P_esc}
P_{xy}^{\rm esc}=\frac{\sigma^{\rm eq}_{xy}}{c_x}.
\ee
The simplest one-dimensional random walk can be described as follows \cite{Doyle}. A man walks along an $n$-block stretch of Madison Avenue. He starts at corner $x$ and, with probability 1/2, walks one block to the right and, with probability 1/2, walks one block to the left; when he comes to the next corner he again randomly chooses his direction along Madison Avenue. He continues until he reaches corner $n+1$, which is home, or corner 1, which is a bar. 

The random walk above would correspond to a one-dimensional circuit (identical resistors connected in series) with the battery connected across nodes $x$ and $n+1$. The escape probabily would be the probability of the man getting home before getting to the bar. Conversely, the probability for the man to reach the bar before getting home would correspond to the equivalent resistance between nodes $1$ and $x$. Note that such a circuit is a partial case of the circuit in Fig. \ref{Fig:InfCirc} for $R_2$ or $\alpha \rightarrow \infty$. Madison Avenue then corresponds to the top line of resistors $R_1$.

For a finite $R_2$, one can think of a random walk along Madison Avenue with a \lq{}\lq{}gutter\rq{}\rq{}: at every corner (other than the bar or home) there is an equal probaility to go left or right  
\be 
p_\leftarrow=p_\rightarrow=\frac{\sigma_1}{2\sigma_1+\sigma_2}\equiv \frac{\alpha}{2\alpha+1}
\ee
 and a non-zero probability to fall into the gutter 
\be
 p_\downarrow =\frac{\sigma_2}{2\sigma_1+\sigma_2}\equiv \frac{1}{2\alpha+1}. 
\ee
Note that there is only one edge coming out of node $1$, hence $c_1 = \sigma_1\equiv 1/R_1$. Combining Eqs. (\ref{P_esc}) and (\ref{REQ}) we obtain 
\be
P^{\rm esc}_{1,k}(k)=2\frac{\left({1+\sqrt{{1+4\alpha}}}\right)^{2k}-\left({1-\sqrt{{1+4\alpha}}}\right)^{2k}}
{\left({1+\sqrt{{1+4\alpha}}}\right)^{2k+1}-\left({1-\sqrt{{1+4\alpha}}}\right)^{2k+1}}
\ee
for the probability of the man randomly walking from the bar toward his home $k$ blocks away to end up in the gutter.


\end{document}